\documentclass[10pt]{article}

\usepackage[T1]{fontenc}
\usepackage[bottom]{footmisc}
\usepackage[margin=1.0in]{geometry}
\usepackage[square,numbers]{natbib}
\usepackage[utf8]{inputenc}
\usepackage{amsfonts}
\usepackage{amsmath}
\usepackage{amssymb}
\usepackage{array}
\usepackage{bibentry}
\usepackage{booktabs}
\usepackage{caption}
\usepackage{enumerate}
\usepackage{fancyref}
\usepackage{float}
\usepackage{graphicx}
\usepackage{hyperref}
\usepackage{xcolor}

\newcolumntype{Y}[1]{>{\centering\let\newline\\\arraybackslash\hspace{0pt}}m{#1}}

\hypersetup{hidelinks}
\hypersetup{colorlinks, citecolor=black, linkcolor=black, filecolor=magenta, urlcolor=blue}

\newcommand{\rqformat}[1]{{\textbf{RQ#1}}}  
\newcommand{\rqstudents}{\rqformat{1} (student use)}
\newcommand{\rqthink}{\rqformat{2} (student thoughts)}
\newcommand{\rqinstructors}{\rqformat{3} (instructor benefit)}

\newcommand{\ritprofs}{
  Rochester Institute of Technology,
  Department of Computer Science,
  Rochester, New York, USA}

\newcommand{\ritstudents}{
  Rochester Institute of Technology,
  Golisano College of Computing and Information Sciences,
  Rochester, New York, USA}

\newcommand{\ur}{
  University of Rochester,
  Center for Professional Development and Education Reform,
  Rochester, New York, USA
}

\begin{document}

\title{Effective Feedback for Introductory CS Theory:\\A JFLAP Extension and Student Persistence}

\author{
  \begin{tabular}{Y{45mm} Y{45mm} Y{45mm}}
    Ivona Bez\'akov\'a\footnote{\ritprofs} &
    Kimberly Fluet\footnote{\ur} &
    Edith Hemaspaandra\footnotemark[1]
  \end{tabular}
  \\ \\
  \begin{tabular}{Y{45mm} Y{45mm}}
    Hannah Miller\footnote{\ritstudents} &
    David E.~Narv\'aez\footnotemark[3]
  \end{tabular}
}

\date{}


\maketitle

\vspace{-5mm}


\begin{abstract}

  Computing   theory  analyzes   abstract   computational  models   to
  rigorously study the computational difficulty of various problems.
  Introductory computing  theory can be challenging  for undergraduate
  students, and  the main  goal of  our research  is to  help students
  learn these computational models.
  The most common  pedagogical tool for interacting  with these models
  is the Java Formal Languages and Automata Package (JFLAP).
  We  developed  a  JFLAP  server extension,  which  accepts  homework
  submissions from  students, evaluates  the submission as  correct or
  incorrect,  and provides  a witness  string when  the submission  is
  incorrect.  Our  extension currently  provides witness  feedback for
  deterministic  finite  automata, nondeterministic  finite  automata,
  regular expressions, context-free grammars, and pushdown automata.

  In Fall  2019, we  ran a preliminary  investigation on  two sections
  (Control   and  Study)   of   the   required  undergraduate   course
  Introduction to Computer Science Theory.
  The  Study section  used our  extension for  five targeted  homework
  questions,  and  the  Control  section solved  and  submitted  these
  problems using traditional means.
  Our results  show that  on these five  questions, the  Study section
  performed better on average than the Control section.
  Moreover, the  Study section persisted in  submitting attempts until
  correct, and from  this finding, our preliminary  conclusion is that
  minimal  (not  detailed  or   grade-based)  witness  feedback  helps
  students to truly learn the concepts.
  We describe  the results that support  this conclusion as well  as a
  related  hypothesis  conjecturing  that with  witness  feedback  and
  unlimited number of submissions,  partial credit is both unnecessary
  and ineffective.

\end{abstract}


\section{Introduction}

Computing theory is difficult for beginner students since the concepts
are  abstract.  In  introductory  theory  courses, students  construct
computational   models   such   as  deterministic   finite   automata,
nondeterministic  finite automata,  regular expressions,  context-free
grammars, and pushdown  automata (DFAs, NFAs, RegExs,  CFGs, and PDAs,
respectively).  The  most popular graphical interface  for students to
interact with  these concepts~\cite{fiftyyearsofautomata} is  the Java
Formal  Languages  and  Automata Package  (JFLAP)  \cite{jflapwebsite,
  jflapbook, jflap2009}.

\begin{figure}[H]
  \centering
  \includegraphics[width=0.4\columnwidth]{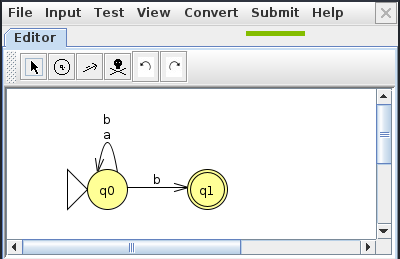}
  \caption{JFLAP with the DAVID extension ``Submit'' option.}
  \label{fig:jflap-david-screenshot}
\end{figure}

For  our   Automated  Feedback   in  Undergraduate   Computing  Theory
\cite{afctgrant}  research,  we  developed  the  Didactic  And  Visual
Interface  for  Development (DAVID)  extension  to  JFLAP.  The  DAVID
extension  accepts  introductory   computer  science  theory  homework
submissions  from students  and sends  each submission  to a  feedback
server, which automatically checks  a student's submission against the
instructor's  correct solution;  the  server  then provides  immediate
feedback  to  the  student  if  the  submission  is  correct  or  not.
\figurename~\ref{fig:jflap-david-screenshot}  shows   JFLAP  with  the
DAVID extension.

In Fall 2019,  we conducted a successful  preliminary investigation of
the beta  version of the  DAVID extension.  In this  investigation, we
compared  a  Control  section  and  a Study  section  of  students  in
Introduction  to  Computer  Science  Theory.  The  Study  section  was
required to use  the DAVID extension to submit  five targeted homework
questions, and we considered  these exploratory and deliberately broad
research questions  about students  and instructors.   Our preliminary
research  questions were  deliberately broad  to prevent  artificially
limiting  ourselves.   We  discuss  our most  interesting  finding  in
Section~\ref{sec:persistence}.
\begin{description}
\item  [\textbf{RQ1}]  How  do   students  use  the  DAVID  extension?
  (Results in Section~\ref{sec:statistical-analysis}.)
\item  [\textbf{RQ2}]   What  were  students'  experiences   with  the
  extension?  (Results in Section~\ref{sec:surveys}.)
\item [\textbf{RQ3}]  How do  instructors benefit from  the extension?
  (Results in Section~\ref{sec:persistence}.)
\end{description}

A student's submission  to the DAVID extension is  either incorrect or
correct.   If a  student's submission  is incorrect,  then the  server
provides  immediate  feedback to  the  student  via a  \textit{witness
  string}, which is a string that the incorrect submission accepts (or
rejects) but the correct solution  should reject (or accept).  For the
models used in introductory theory,  these witness strings are usually
just a  few symbols  long.  Naturally, if  the submission  is correct,
then the extension reports a  simple ``Correct!''  to the student.  We
call this feedback of a  witness string \textit{witness feedback}, and
we found  that when  given only witness  feedback, students  tended to
persist in submitting attempts until correct.

Our  preliminary   investigation  had  multiple   promising  outcomes,
including verifying the  set-up of the DAVID  extension, learning what
additional  telemetry would  be valuable  for our  full investigation,
understanding the  practical workflow to  make the extension  easy for
instructors  to use,  and  analyzing the  collected  data of  homework
submissions, student surveys, and students' grades.

\section{Relevant Literature}\label{sec:related-work}

Norton~\cite{daphne}     studied    near-linear     time    algorithms
\cite{hopcroft1971,gries1972}   for   DFA    equivalence   and   wrote
JFLAP-compatible  code  to prove  DFA  equivalence  and to  produce  a
witness string if two DFAs  are not equivalent.  Since CFG equivalence
is an undecidable problem, Sorrell used CFGAnalyzer \cite{cfganalyzer}
to experimentally show  that checking all strings up  to length $k=10$
suffices for  typical homework  assignments.  Sorrell's  work included
integrating the PicoSAT solver~\cite{biere2008} into CFGAnalyzer; this
work  is  the CFGSolver  software  \cite{sorrell}  used by  the  DAVID
extension.  To be conservative, our extension checks all strings up to
length $k=15$ for strings generated by a CFG submission.  We announced
the setup of our preliminary investigation in~\cite{sigcse2020}.

Automata Tutor is  an alternative feedback tool  for computing theory.
Version 2  \cite{websiteAT2} provides a graphical  user interface with
roughly  the functionality  of JFLAP  as well  as additional  features
related  to  checking  equivalence for  regular  languages,  including
witness  feedback   and  automated  grading   of  DFAs~\cite{alur2013,
  dantoni2015}.  Presented in May  2020, Version 3~\cite{AT3paper} has
many added  features, including  those that we  implemented on  top of
JFLAP.

While  the  tools  are  similar,  the  focus  of  the  Automata  Tutor
study~\cite{AT3paper}  and  our   investigation  are  very  different.
Automata Tutor  focuses on automation, including  partial grading.  We
focus on  educational research, attempting to  measure the educational
benefits of the  DAVID extension, which provides a  witness string for
incorrect  solutions and  does not  venture into  partial credit.   In
fact, the most interesting finding of our preliminary investigation is
that the majority  of students persisted in  submitting attempts until
the DAVID extension reported correct.   We discuss the implications of
the persistence result in Section~\ref{sec:persistence}, from which we
hypothesize  (Section~\ref{sec:hypotheses}) that  witness feedback  is
the appropriate type  of feedback (Section~\ref{sec:witness-feedback})
and   that   partial   credit   is   not   needed   in   our   setting
(Section~\ref{sec:no-partial-credit}).

For this paper, we use the term \textit{feedback}, which is defined as
``information provided by an agent (e.g., teacher, peer, book, parent,
experience) regarding aspects of  one's performance or understanding''
\cite{hattie2007}.
In particular, we are interested in \textit{intermediate feedback}
(Section~\ref{sec:witness-feedback}), which is a type of
\textit{formative feedback}~\cite{sadler1989}.


\section{Fall 2019 Preliminary Investigation}\label{sec:prelim-eval}

In Summer  2019, we developed the  DAVID extension.  In Fall  2019, we
ran a preliminary investigation  on approximately 65 students enrolled
in Introduction  to Computer  Science Theory, and  in Spring  2020, we
analyzed   the  investigation   results.    These   results  and   the
implications for our future work are discussed in this paper.

For  our Fall  2019 preliminary  study with  the DAVID  extension, two
synchronized sections  of the Introduction to  Computer Science Theory
course  shared the  same instructor,  course delivery  style, homework
assignments, similar midterm exams, and  an identical final exam.  The
majority  of students  ($>80\%$)  in both  sections  were majoring  in
Computer Science; the second-most popular major ($<15\%$) was Software
Engineering.

For both sections, we collected  homework and grade data, surveyed the
students about  their experiences in  the course, and asked  the Study
section additional  survey questions about their  experiences with the
DAVID extension, their perception of the extension, and their thoughts
on automated feedback in general.  We monitored how the Study students
used the DAVID extension.  We also interviewed the instructor.  At the
end of the  course, there were 35 students in  the Control section and
29~students in the Study section who gave consent for their data to be
used in our investigation.  There were eleven homework assignments for
the semester with an average of five questions per homework.

Out of these  eleven assignments, we compared  the performance between
the two  sections on five targeted  homework questions (one each  on a
DFA,  NFA, RegEx,  CFG, and  PDA),  which were  identical between  the
sections.   For these  five targeted  homework questions,  the Control
section  submitted their  homework solutions  to these  five questions
using  traditional means  the Study  section was  required to  use the
DAVID extension.
The text  of the five targeted  homework questions is below.   The CFG
question, which was especially challenging, had by far the most number
of submissions to the DAVID extension.

\begin{enumerate}

\item \textit{DFA.} Draw the state  diagram of a finite automaton that
  accepts the language of all strings  over $\{a, b\}$ that contain at
  least  2 $b$'s  and do  not contain  the substring  $bb$.  In  other
  words,  a string  is accepted  only  if both  conditions hold.  Your
  finite automaton should not be overly complicated.

\item \textit{NFA.}   Draw the state  diagram of an NFA  accepting the
  language of  all strings over  $\{a, b\}$  that either start  or end
  with  the   substring  $aba$.   For   full  credit,  you   must  use
  nondeterminism where possible  to make your state  diagram as simple
  as possible.

\item \textit{RegEx.}  Give  a regular expression for  the language of
  all strings over $\{0, 1\}$ that  have neither the substring 000 nor
  the  substring  111.  In  other  words,  the language  contains  all
  strings  for  which neither  symbol  ever  appears more  than  twice
  consecutively.

\item \textit{CFG.}   Give a  CFG that generates  the language  of all
  strings  over $\{0,  1\}$  that  have more  consecutive  0's at  the
  beginning  of the  string than  consecutive 1's  at the  end of  the
  string.   (For  example,  the  following  strings  are  all  in  the
  language: \{0, 001,  00001010101010111, 0111111110\}.  The following
  strings are  all not  in the language:  \{$\epsilon$, 01,  10, 0011,
  0010000000111\}.)

\item \textit{PDA.}  Let
  $L  =  \{w \in  \{a,  b\}^{*}  |~w\text{  has more  }a\text{'s  than
  }b\text{'s}\}$.  Draw  the state diagram  of a PDA that  accepts the
  language $L$.  Your PDA should not be overly complicated.

\end{enumerate}


\section{Results}

\subsection{Statistical analysis}\label{sec:statistical-analysis}

\begin{figure}[H]
  \centering
  \includegraphics[width=252pt]{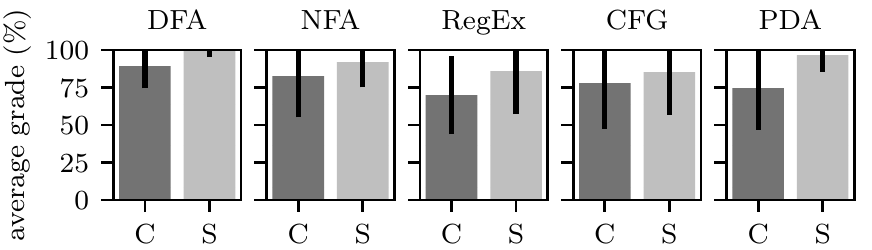}
  \caption[Homework grade average and  standard deviation for the five
  targeted homework questions.]
  {Homework grade average and standard deviation for the five targeted
    homework questions.   ``C'' is the  Control section; ``S''  is the
    Study  section.   The  $y$-axis  limits are  the  same  among  the
    subplots.}
  \label{fig:hw-avg-stdev}
\end{figure}

For \rqstudents, for both the final  exam grade and the overall course
grade, the  Study section scored  lower than the Control  section, and
both     lower      scores     were      statistically     significant
(Table~\ref{tab:overall-grades});  these differences  are  not due  to
chance, which tells  us that the two sections had  different levels of
academic strength.   However, on the five  targeted homework questions
where the  Study section  used the DAVID  extension for  feedback, the
Study  section's average  grade  was always  higher  than the  Control
section's average grade (\figurename~\ref{fig:hw-avg-stdev}).

\begin{figure}[H]
  \centering
  \includegraphics[width=252pt]{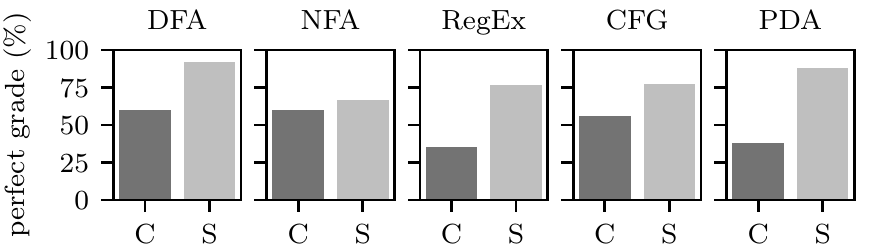}
  \caption[The percentage  of students who  earned a perfect  grade on
    the five targeted questions  from an experienced professor.]
    {The percentage of students who earned a perfect grade on the five
      targeted questions from an  experienced professor.  ``C'' is the
      Control  section;  ``S'' is  the  Study  section.  The  $y$-axis
      limits are the same among the subplots.}
  \label{fig:correct-grade-normalized}
\end{figure}

The  Study  section strongly  outperformed  the  Control section  with
respect to  the percent  of perfect homework  grades for  the targeted
homework  questions  (\figurename~\ref{fig:correct-grade-normalized}).
(The  Study  section NFA  percentage  is  low  compared to  the  other
questions because the NFA submissions  through the DAVID extension are
not  checked  for  ``sufficient''  nondeterminism,  which  means  that
submissions counted  as correct by  the extension did  not necessarily
earn a perfect grade.)  We saw high engagement with the extension from
the  Study  students:  on  average, students  submitted  9  times  per
homework question.

We  used  ANOVA  with a  threshold  of  $p  <  0.050$ to  compare  the
differences  between  the  Control  section  and  the  Study  section.
Table~\ref{tab:stats} summarizes the statistical results.  On three of
the five targeted homework questions, the Study section's higher score
was  statistically significant;  for the  other two  targeted homework
questions, there  was no statistically significant  difference between
the  sections.   Because  the  Study  section  was  academically  less
proficient  than the  Control section,  it is  particularly noteworthy
that the  Study section scored  significantly higher than  the Control
section on the DFA, RegEx, and PDA questions.

\begin{table}[H]
  \centering
  \begin{tabular}{lcc}
    \toprule
                       & $p$ value & Significance                     \\
    \midrule
    DFA                & 0.001     & Study scored higher than Control \\
    NFA                & --        & no statistical difference        \\
    RegEx              & 0.030     & Study scored higher than Control \\
    CFG                & --        & no statistical difference        \\
    PDA                & 0.000     & Study scored higher than Control \\
    Final exam grade   & 0.014     & Study scored lower than Control  \\
    Course grade       & 0.011     & Study scored lower than Control  \\
    \bottomrule
  \end{tabular}
  \caption[Statistical results.]
  {Statistical results.  A dash indicates no statistically significant
    $p$ value.  All analysis used a threshold of $p < 0.050$.}
   \label{tab:stats}
\end{table}


To determine the  relative academic strength of  the Control vs.~Study
sections, we used the final exam  grade, the overall course grade, and
the  instructor's perception.
The final exam was identical for both sections, and for the final exam
grade,
the Control section mean was $79.8\% \pm 17.0\%$ for 35 students,
and the Study section mean was $70.3\% \pm 12.9\%$ for 29 students.
This  difference  was  statistically  significant ($p  =  0.014$)  and
therefore  is  not  due  to  chance.  For  the  overall  course  grade
(Table~\ref{tab:overall-grades}), the Control section performed better
than the Study section with respective means of $84.9\% \pm 9.9\%$ and
$78.0\%   \pm    11.0\%$,   which   was    statistically   significant
($p = 0.011$).

\begin{table}[H]
  \centering
  \begin{tabular}{c crr ccccc}
    \toprule
             &       &       &        & \multicolumn{5}{c}{Quartiles}   \\
                                        \cmidrule(lr){5-9}
             & $n$ &  \multicolumn{1}{c}{$\mu$} &  \multicolumn{1}{c}{$\sigma$} &  min & 25\% & 50\% & 75\% & max \\
    \midrule
    Control  &    35 &  84.9 &  9.9 & 54.3 & 80.0 & 86.3 & 92.9 & 96.7  \\
    Study    &    29 &  78.0 & 11.0 & 39.0 & 76.0 & 78.3 & 86.4 & 91.5  \\
    \bottomrule
  \end{tabular}
  \caption[Overall  course grades  for both  sections.]
    {Overall  course grades  for both  sections.  The  number of
    students  is  $n$, the  mean  grade  is  $\mu$, and  the  standard
    deviation of the grades is $\sigma$.}
  \label{tab:overall-grades}
\end{table}

Additionally, the course  instructor said, ``I do  think it's probably
correct  that just  kind of  as  an average  performance, the  Control
section  was  a  little  sharper [than  the  Study  section].''   This
evidence of the final exam and the overall course grade as well as the
instructor's perception supports our claim  that the Study section was
academically less successful than the Control section.  Therefore, the
statistically significant higher scores by  the Study section on three
of the five  targeted homework questions are even  more striking since
they outperformed the academically stronger Control section.


\subsection{Survey responses}\label{sec:surveys}

For \rqthink, we surveyed students  in both sections twice about their
experiences  in  the  course.   On  each  survey,  the  Study  section
responded  positively  about  the   DAVID  extension.   We  share  two
representative free-text responses: on Survey~1 (after DFAs and NFAs),
Student \#21 said, ``[The] DAVID extension is a very helpful tool.  It
helped me understand DFA and NFA  better.''  On Survey~2 (near the end
of  the course),  Student \#3  said, ``I  love the  idea of  the DAVID
extension, but it could use  improvement.  Definitely don't get rid of
it.''

We     report      the     survey     response      percentages     in
Table~\ref{tab:survey-percentages}  and  Table~\ref{tab:resub},  where
the  column  names  are   ``strongly  agree  +  agree,''  ``neutral,''
``disagree +  strongly disagree.''  Table~\ref{tab:survey-percentages}
shows the Study section responses to the two survey questions below.
\begin{description}
\item [Q1] The DAVID extension helped  me solve the \{DFA, NFA, RegEx,
  CFG, PDA\} questions, and
\item [Q2]  The DAVID extension  helped me understand the  \{DFA, NFA,
  RegEx, CFG, PDA\} questions.
\end{description}

From the survey  responses, the Study students did feel  helped by the
DAVID extension  (Table~\ref{tab:survey-percentages}).  For  survey Q1
(DAVID extension helped the student solve), on the DFA, RegEx, and PDA
questions where the Study section did better than the Control section,
the  majority of  Study students  strongly agreed  or agreed  that the
extension  helped them  to  solve that  problem  (54.9\%, 60.7\%,  and
42.9\%,  respectively).  For  survey  Q2 (DAVID  extension helped  the
student understand), the majority of Study students strongly agreed or
agreed that the extension helped  them understand the concepts (41.9\%
for DFA and  46.4\% for RegEx).  For PDAs, the  percentage of students
who responded neutral  (42.9\%) was near equal to  those who responded
strongly agree or agree (39.3\%); however, the percent of students who
strongly agreed or agreed was more than double those that disagreed or
strongly disagreed (17.8\%).

\begin{table}[H]
  \centering
  \begin{tabular}{l ccc ccc}
    \toprule
          & \multicolumn{3}{c}{Q1: Solve} & \multicolumn{3}{c}{Q2: Understand} \\
            \cmidrule(lr){2-4}         \cmidrule(lr){5-7}
          &  SA+A            &  N      &  D+SD   &  SA+A            &  N      &  D+SD   \\
    \midrule
    DFA   &  \textbf{54.9\%} &  19.4\% &  25.7\% &  \textbf{41.9\%} &  29.0\% &  29.1\% \\
    NFA   &  51.6\%          &  22.6\% &  25.8\% &  45.2\%          &  25.8\% &  29.0\% \\
    RegEx &  \textbf{60.7\%} &  21.4\% &  17.9\% &  \textbf{46.4\%} &  32.1\% &  21.5\% \\
    CFG   &  39.3\%          &  42.9\% &  17.8\% &  42.8\%          &  35.7\% &  21.5\% \\
    PDA   &  \textbf{42.9\%} &  39.3\% &  17.8\% &  \textbf{39.3\%} &  42.9\% &  17.8\% \\
    \bottomrule
  \end{tabular}
  \caption[Study section survey responses.]
  {Study  section survey  responses.   Bold  percentages indicate  the
    homework questions where the Study section performed statistically
    better than the Control section.}
  \label{tab:survey-percentages}
\end{table}

On  the  second  survey,  we  also  asked  the  Study  students  about
resubmission and partial credit.  The  survey questions are below, and
the Study students' responses are in Table~\ref{tab:resub}.
\begin{description}
\item [Q3]  The DAVID extension  should allow users to  resubmit until
  correct.
\item [Q4]  Assignments submitted  via the  DAVID extension  should be
  graded to allow partial credit.
\item [Q5]  Assignments submitted  via the  DAVID extension  should be
  graded as correct/incorrect (i.e., without partial credit).
\end{description}

Students  overwhelmingly agreed  ($>95\%$)  that  the DAVID  extension
should allow  users to  resubmit until correct,  and they  also agreed
that assignments  should be graded  to allow partial credit,  but they
disagreed that  assignments should  be graded  as strictly  correct or
incorrect.    In   Section~\ref{sec:persistence},   we   discuss   the
implications of  these results  for student  learning and  for optimum
design of homework feedback.

\begin{table}[H]
  \centering
  \begin{tabular}{ l rrr}
    \toprule
             & \multicolumn{1}{c}{SA+A} & \multicolumn{1}{c}{N} & \multicolumn{1}{c}{D+SD} \\
    \midrule
    Q3: Allow resubmit       &  \textbf{96.4\%}  &   3.6\%  &   0.0\% \\
    Q4: Allow partial credit &  \textbf{71.4\%}  &  21.4\%  &   7.1\% \\
    Q5: No partial credit    &  10.7\%           &  28.6\%  &  \textbf{60.7\%} \\
    \bottomrule
  \end{tabular}
  \caption[Survey response  percentages from 28 students  in the Study
    section  about resubmission  and partial  credit.]
    {Survey response  percentages from 28 students  in the Study
    section  about resubmission  and partial  credit.  Bold  shows the
    highest percentage response for each question.}
  \label{tab:resub}
\end{table}


\subsection{The phenomenon of persistence}\label{sec:persistence}

For \rqinstructors, we  saw that with the immediate  feedback from the
DAVID extension,  more students  in the  Study section  did eventually
solve         the          homework         problems         correctly
(\figurename~\ref{fig:correct-grade-normalized}),  which  benefits  an
instructor because  grading correct  submissions is faster  and easier
than grading incorrect submissions.  We have named this benefit of our
extension \textit{grading triage}.

\begin{figure}[H]
  \centering
  \includegraphics[width=252pt]{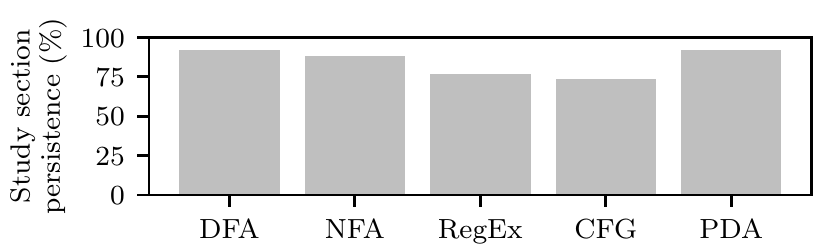}
  \caption{Persistence of the Study  section until the DAVID extension
    reported ``Correct.''}
  \label{fig:study-persistence}
\end{figure}

Recall  that the  focus of  our work  is about  developing a  homework
feedback  server for  students, and  our work  is not  about automatic
grading.   In  \figurename~\ref{fig:study-persistence},   we  see  the
percentage  of students  who continued  to submit  attempts until  the
extension reported  ``Correct.''\footnote{In this figure, we  omit the
  occasional student  who did not  get meaningful feedback  because of
  syntax  errors;  for  example,  mismatched  parentheses  in  regular
  expressions did  not display an  error message to the  student.}  We
call this behavioral phenomenon \textit{persistence}.

Since  the DAVID  extension does  not grade  submissions, a  student's
homework grade must  be manually assigned by an instructor  (or a TA).
While  grading homework  which has  received feedback  from the  DAVID
extension, an instructor  can easily look at  the corresponding output
from  the DAVID  extension.   For incorrect  submissions, the  witness
string feedback is  a valuable resource to help an  instructor see the
error in  the submission.  Furthermore, even  for correct submissions,
this  grading triage  benefit is  useful for  regular expressions  and
CFGs,  where  manual  verification  of   a  correct  solution  can  be
complicated.

\begin{figure}[H]
  \centering
  \includegraphics[width=0.7\columnwidth]{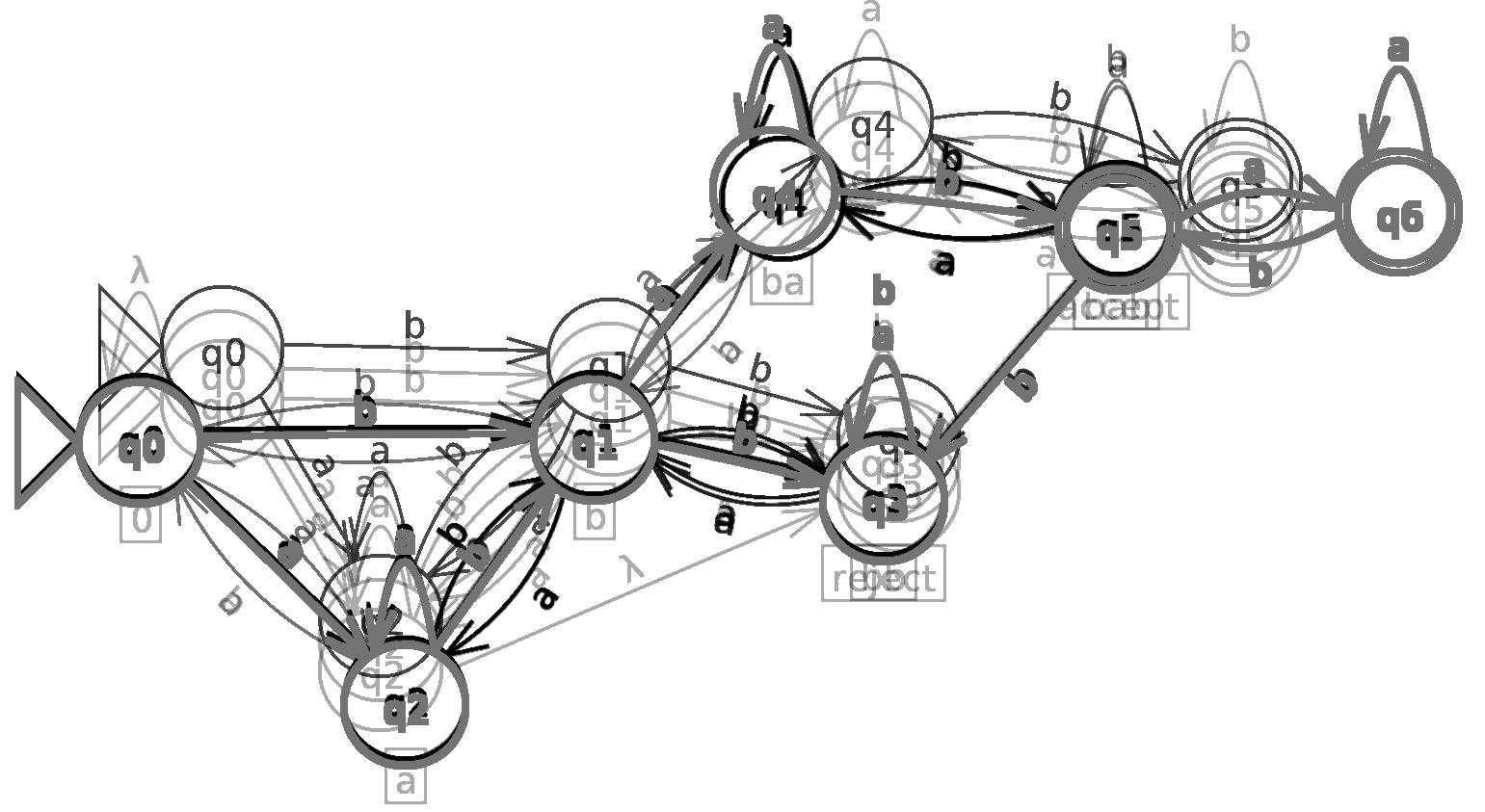}
  \caption[One student's 26 unique submissions for a DFA.]
    {One  student's 26  unique  submissions for  a  DFA for  the
    homework question: ``Draw  a DFA that accepts the  language of all
    strings over $\{a,  b\}$ that contain at least 2  $b$'s and do not
    contain  the substring  $bb$.''   The thin  gray  lines are  fewer
    similar  submissions,   and  the  gray  lines   are  many  similar
    submissions.   The thick,  dark  gray line  is  the correct  final
    submission.}
  \label{fig:dfa-26-avg-of-all-inputs}
\end{figure}

In fact,  this behavior is  voluntary persistence since  students knew
they  would  get  partial  credit for  incorrect  solutions,  and  yet
students  still persisted  until  getting a  correct  answer from  the
extension.   \figurename~\ref{fig:dfa-26-avg-of-all-inputs}   shows  a
single student's persistence  with our extension for  the DFA homework
question in Section~\ref{sec:prelim-eval}.  We expect that persistence
will   be  even   higher  if   there   is  no   partial  credit.    In
Section~\ref{sec:discussion}, we  discuss the phenomenon  of voluntary
persistence as  a key insight  for intermediate feedback  on incorrect
submissions   (Section~\ref{sec:witness-feedback})  and   for  partial
credit (Section~\ref{sec:no-partial-credit}).

\section{Discussion}\label{sec:discussion}


\subsection{Hypotheses for our full investigation}\label{sec:hypotheses}

For our exploratory and preliminary Fall 2019 investigation, our
research questions were deliberately broad so that we could find
interesting directions for our future work.  Our most interesting
result was student persistence: with only the short witness string as
feedback, students persisted in submitting attempts until the DAVID
extension reported ``Correct''
(\figurename~\ref{fig:study-persistence}).  Our preliminary
investigation led us to these hypotheses for our full investigation.
\begin{description}
\item  [\textbf{H1}]   In  our   setting,  witness  feedback   is  the
  appropriate type of feedback.
\item [\textbf{H2}] In our setting, partial credit is not needed.
\end{description}


\subsection{The case for witness feedback}\label{sec:witness-feedback}

Witness  feedback is  the \textit{minimal  reasonable feedback}.   The
educational literature calls this \textit{minimal intervention}, which
has been found to promote  better learning than more detailed feedback
\cite{wiliam1999}.   Minimal  reasonable feedback  convincingly  shows
that the submission is wrong, but the feedback does not give any hints
for  fixing  the  submission.   Creighton  \textit{et  al.}~say,  ``If
feedback attempts to provide too  much guidance, there is nothing left
for the student to do  or learn'' \cite{creighton2015}.  Similar ideas
are also in \cite{day1993,rakoczy2019}.

Witness feedback,  which is very  minimal, gives the student  a reason
why a submission is wrong, but  the witness feedback does not tell the
student how to correct the  mistakes.  Thus, witness feedback will not
lead  a student  to the  correct solution;  instead, the  student must
think independently.  In  fact, this sort of  minimal witness feedback
mimics the  feedback that  an instructor  or a tutor  would give  to a
student seeking help by providing a short witness string showing where
the  student's attempt  is  incorrect.  Feedback  of  a single,  short
witness  string requires  the student  to actively  learn in  order to
solve the question.

One problem  with allowing students  to submit  as many times  as they
like is that students may try  to random-walk to the correct solution.
Because the witness does not give  information about how to change the
submission,  the  potential  of  randomly converging  on  the  correct
solution is not an issue.  For example,
\figurename~\ref{fig:dfa-26-avg-of-all-inputs}  shows  a  student  who
clearly has the  right idea and is refining the  solution based on the
minimal feedback of the DAVID extension.

However,  when giving  more  detailed  feedback, encouraging  ``random
walks to a solution''  can be an issue.  In addition,  there is a real
risk  of ``over  helping'' and  leading  the student  to the  solution
step-by-step in  such a way  that the student contributes  very little
(even  though  the  student  may not  realize  this).   Finally,  more
detailed  feedback may  encourage students  to make  local fixes  that
create more and more bloated submissions.

As  a related  but distinct  point, giving  more detailed  feedback is
hard.   For example,  if there  is  more than  one way  to approach  a
problem, feedback can easily steer students into a direction that does
not  correlate to  the  student's approach.   For  a simple,  standard
example,  there are  two  ways to  approach designing  a  CFG for  the
language $\{a^ib^k \ |  \ i \neq k\}$.  The first one  is to cross off
$a$'s and $b$'s until you are left with just $a$'s or just $b$'s. This
corresponds to a CFG with the following rules.
  \begin{eqnarray*}
  S & \rightarrow & aSb \ | \ A \ | \ B\\
  A & \rightarrow & aA \ | \ a\\
  B & \rightarrow & bB \ | \ b
  \end{eqnarray*}
  The  second one is to  view this language as  the union of
  two  cases: ``$i  < k$''  and ``$i  > k$.'' This corresponds to a CFG with the following rules.
  \begin{eqnarray*}
  S & \rightarrow & A \ | \ B\\
  A & \rightarrow & aAb \ | \ aA \ | \ a\\
  B & \rightarrow & aBb \ | \ bB \ | \ b
  \end{eqnarray*}
Telling a  student who is following  the first approach to  ``think of
the  language as  a union''  is  not helpful  at best.
Of  course, an  instructor could  add  both approaches  to a  feedback
system,  and  for  regular   expressions,  Automata  Tutor  uses  this
technique of adding multiple  ``reasonable'' approaches to the grading
system.
In general,  it will  be hard for  an instructor to  come up  with all
reasonable  approaches, and  it seems  impossible to  do this  with an
automated program.

\subsection{The case against partial credit}\label{sec:no-partial-credit}

Our  investigation  focuses  on automated  feedback  for  intermediate
student  submissions.  Of  course, grading  is a  form of  feedback as
well.  Automata Tutor uses the current grade as (part of) the feedback
on  intermediate student  submissions.   Automated grading  is a  very
interesting  topic in  its  own right,  particularly  given the  large
numbers of CS majors.

For  incorrect attempts,  using partial  grade credit  as feedback  in
addition  to  the  witness  suffers from  the  problems  described  in
Section~\ref{sec:witness-feedback}.   If  students  are  chasing  more
partial credit,  then they may be  randomly trying to converge  on the
wrong thing  (more points)  rather than the  right thing  (the correct
language).   In \cite{cain2016},~Cain and  Babar paraphrase  (Skinner
2014), saying,  ``Attaching marks  to an  assessment task  means that,
from the  student's perspective, the  task will play a  summative role
and feedback is not seen as formative.''
They continue, ``Interestingly, it has been reported that students pay
more  careful  attention to  feedback  when  there are  no  associated
marks~\cite{black1998}  or put  another  way  `marks' reduced  student
attention to formative feedback.''
In other words,  it is difficult to design a  good scoring system that
really drives to the correct solution.

Automated partial credit has other problems and drawbacks.  One
obvious problem is that different instructors may want to assign
different amounts of partial credit; although in practice, instructors
would probably accept ``reasonable'' partial credit (for example, no
grade ``inversions,'' meaning that better solutions should not get
less credit).

However,  it  is  hard  if  not  impossible  to  automatically  assign
reasonable  partial  credit.   For  example, in  Automata  Tutor,  the
fraction of  points assigned for a  CFG is computed as  an estimate of
$\frac{|A \cap B|}{|A  \cup B|}$, where $A$ is  the language generated
by the submitted CFG and $B$ is the correct language.
If we consider a  simple language like $\{a^{2i}b^i \ |  \ i \geq 1\}$
with the standard solution
\begin{equation*}
  S \rightarrow aaSb \ | \ aab,
\end{equation*}
then the  four solutions in Table~\ref{tab:four-close},  which are all
fairly close, get no credit at all!
This  is  not  meant  to   be  negative  about  Automata  Tutor; any
language-based partial credit metric will have similar problems.

\begin{table}[H]
  \centering
  \begin{tabular}{ll}
  \toprule
  \multicolumn{1}{c}{CFG} & \multicolumn{1}{c}{language} \\
  \midrule
  $S \rightarrow aSbb \ | \ abb$ & $\{a^ib^{2i} \ | \ i \geq 1\}$\\
  $S \rightarrow bbSa \ | \ bba$ & $\{b^ia^{2i} \ | \ i \geq 1\}$\\
  $S \rightarrow aSb \ | \ ab$ & $\{a^{i}b^{i} \ | \ i \geq 1\}$\\
  $S \rightarrow aaSb \ | \ ab$ & $\{a^{2i+1}b^{i+1} \ | \ i \geq 0\}$\\
  \bottomrule
  \end{tabular}
  \caption{Solutions     that     are     close    to     the     rule
    $S \rightarrow  aaSb \ |  \ aab$, but  do not intersect  with that
    language at all.}
  \label{tab:four-close}
\end{table}

Indeed,  Automata  Tutor~\cite{AT3paper}  for   RegExs  looks  at  the
distance from a few ``sensible'' RegExs supplied by the instructor (if
the  submission is  correct,  the student  always  gets full  credit),
stating that  ``...  This  is preferable  to comparing  the languages,
because a  small careless mistake in  the RE [RegEx] can  have a large
impact on the  language.''  We agree (and of course  the same argument
holds for CFGs as well), but it  may be hard for an instructor to list
all  ``sensible'' RegExs,  particularly for  complicated RegExs.   For a simple
example, if a student writes
\begin{equation*}
  a^* + b^* + (a + b)^*
\end{equation*}
instead of
\begin{equation*}
  (a+b)^+,
\end{equation*}
(where  * is  the  Kleene star,  the  infix operator  +  is the  union
operator, and the raised + is the Kleene plus),
then the student will lose a lot of points, even though the submission
is only missing the empty string.

On a final note, students and instructors may feel that not giving
partial credit is overly harsh.  Indeed, when we asked our students on
the survey (Table~\ref{tab:resub}), they were overwhelmingly ($>95\%$)
in favor of the DAVID extension giving partial credit.  However, there
are other ways to give students partial credit.  For example, we can
give five CFG questions and ask the students to submit four, which is
a stress-decreasing approach that works well in many situations,
including exams.
With our  approach of minimal  reasonable feedback, not only  will the
students have  an unlimited number  of retries with  immediate witness
feedback,  but  also  they  can  seek  help  from  the  instructor  or
tutors.


\section{Conclusion}

The DAVID extension is successfully providing feedback for DFAs, NFAs,
RegExs, CFGs, and PDAs.  We  have promising initial results: the Study
section performed better on the  five targeted homework questions than
the Control  section (\figurename~\ref{fig:hw-avg-stdev}).   The Study
students   persisted  (\figurename~\ref{fig:study-persistence})   from
which we conjecture that witness feedback is the right feedback.

Our future  work will continue  the educational focus  of \rqstudents,
\rqthink, and  \rqinstructors.  Since our Fall  2019 investigation was
preliminary  and  our  extension  targeted  $<10\%$  of  all  homework
questions in the  course, we did not  see (and did not  expect to see)
knowledge transfer as measured by the students' performance on related
but unfamiliar  questions on exams.   In our full  investigation where
students will use our extension  on more homework questions, we expect
to see knowledge transfer.

We are most excited about our unexpected result of student persistence.
We  believe that  as  students  persist in  solving  problems via  our
extension   (\textbf{RQ1}),   students   will   learn   the   material
(\textbf{RQ2}), which  benefits not only  the students but  also their
instructors  (\textbf{RQ3}).
From  our  finding  of  student   persistence,  we  will  examine  our
additional  hypotheses about  witness feedback  and partial  credit as
discussed in Section~\ref{sec:hypotheses}:
\begin{description}
\item  [\textbf{H1}]   In  our   setting,  witness  feedback   is  the
  appropriate type of feedback.
\item [\textbf{H2}] In our setting, partial credit is not needed.
\end{description}

As we prepare for our full  investigation, we look forward to studying
our  preliminary  conclusion that  minimal  witness  feedback is  both
necessary and sufficient for students to learn effectively.


\section*{Acknowledgments}

We thank the SIGCSE 2021 anonymous referees for helpful comments.
We thank Aaron Deever and his students for participating.
We thank our advisory board: Douglas Baldwin, Joan Lucas, and Susan
Rodger.
Research supported in part by NSF grant DUE-1819546.


\bibliographystyle{ACM-Reference-Format}
\bibliography{sigcse-2021}

\end{document}